\documentclass{article}
\usepackage{ijcai24}

\usepackage{times}
\usepackage{soul}
\usepackage{url}
\usepackage[hidelinks]{hyperref}
\usepackage[utf8]{inputenc}
\usepackage[small]{caption}
\usepackage{graphicx}
\usepackage{amsmath}
\usepackage{amsthm}
\usepackage{booktabs}
\usepackage{algorithm}
\usepackage{algorithmic}
\usepackage[switch]{lineno}
\usepackage{multirow} 
\usepackage{url}
\usepackage{tikz}
\usepackage{placeins}

\title{A Survey of Retrieval Algorithms in Ad and Content Recommendation Systems}

\author{
 Yu Zhao 
\and Fang Liu
\affiliations
University of Toronto, 
American Express Company 
\emails
\small{
yzqr.zhao@rotman.utoronto.ca, 
fang.liu@aexp.com }
}
    
\begin{document}

\maketitle

\begin{abstract}
This survey examines the most effective retrieval algorithms utilized in ad recommendation and content recommendation systems. Ad targeting algorithms rely on detailed user profiles and behavioral data to deliver personalized advertisements, thereby driving revenue through targeted placements. Conversely, organic retrieval systems aim to improve user experience by recommending content that matches user preferences. This paper compares these two applications and explains the most effective methods employed in each.

\end{abstract}

\section{Introduction}
In the age of digital information, the ability to deliver personalized organic content and advertisements has become a cornerstone of user engagement and revenue generation strategies. Ad recommendation and content recommendation systems, powered by sophisticated retrieval algorithms, are at the forefront of this endeavor. Ad targeting aims to maximize user engagement and conversion rates by delivering highly personalized advertisements based on detailed user profiles and behavioral data. On the other hand, organic retrieval systems enhance user experience by suggesting content that aligns with users’ interests and preferences.

This paper provides a brief survey of retrieval algorithms used in ad recommendation and content recommendation systems. It explores the underlying mechanisms of these algorithms, including content-based filtering, collaborative filtering, and hybrid systems. The two-tower model, a prominent deep learning architecture in recommendation systems, is examined in detail, along with its training, inference, and retrieval processes. Additionally, the paper discusses the challenges and considerations associated with these systems, such as the cold start problem, data quality, and privacy concerns. By comparing ad targeting and organic retrieval, the survey aims to shed light on their unique objectives, data utilization methods, and impact on user experience. 

While we do not explore this idea further in this paper, it is worth noting that the retrieval methods discussed bear a strong resemblance to Retrieval-Augmented Generation (RAG) \cite{hu2024rag} for large language models (LLMs). In RAG, a retrieval component is used to fetch relevant documents or data points, which are then used to inform and enhance the generation process of the LLM. Similarly, in recommendation systems, retrieval methods serve to bring forth pertinent user and item data, which can then be leveraged by the recommendation algorithms to produce highly personalized and accurate recommendations. This parallel highlights the potential for cross-pollination of techniques between retrieval-based recommendation systems and LLM-enhanced retrieval-augmented generation models.

\section{Ad Targeting Models}
Ad targeting is a vital tool for modern digital marketing, aiming to deliver personalized advertisements to specific audiences to maximize engagement and conversion rates. One of the key methodologies behind ad targeting is the use of machine learning and inverted index, a data structure widely used in information retrieval systems to efficiently match user profiles with relevant ads. We will explore the concept of ad targeting in detail, including its implementation through and various targeting strategies such as age, gender, re-targeting, keyword targeting, behavioral targeting, and contextual targeting. \cite{yan2009much} proved the effectiveness of behavior targeting through controlled experimentation. 

\subsection{Inverted Index}
Ad targeting leverages the inverted index \cite{scholer2002compression} to enhance the efficiency and accuracy of delivering personalized advertisements. An inverted index is a data structure that maps content (such as ads) to its associated keywords or attributes, enabling fast and efficient search and retrieval operations.

\begin{figure}[h]
    \centering
    \includegraphics[width=\linewidth]{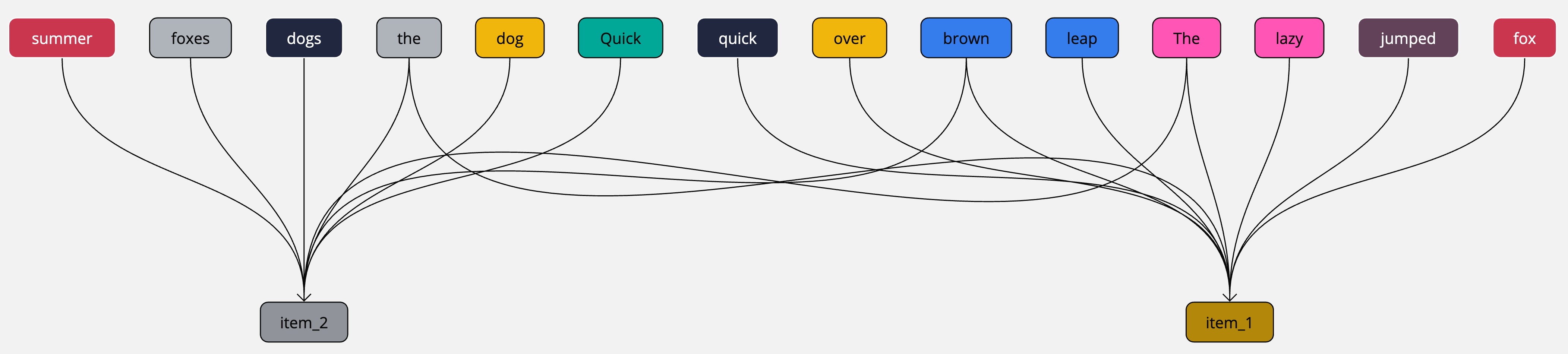}
    \caption{Inverted Index}
    \label{fig:iidx}
\end{figure}
\FloatBarrier

In the context of ad targeting, an inverted index works as follows:
\begin{enumerate}
\item \textbf{Index Creation}: The first step involves creating an index from the available ads. Each ad is broken down into a set of keywords or attributes that describe its content and target audience. These keywords can include demographic information, interests, behaviors, and contextual relevance.

\item \textbf{User Profiling}: Users are profiled based on their online activities, such as search queries, browsing history, social media interactions, and purchase behavior. Each user's profile is similarly decomposed into a set of keywords or attributes that represent their interests and preferences.

\item \textbf{Matching Process}: When a user visits a website or engages with a platform, the ad targeting system retrieves the user's profile and matches it against the inverted index. This process involves looking up the keywords associated with the user in the index to find ads that have matching or closely related keywords.

\end{enumerate}

This approach allows for real-time ad targeting, ensuring that users see ads that are most relevant to their current interests and needs. The inverted index structure is highly efficient, enabling quick lookup and retrieval operations, even for large datasets.

\subsection{Age Targeting}
Age targeting involves delivering ads to users based on their age group. This strategy is particularly effective for products and services that are age-specific, such as toys for children, educational resources for teenagers, or retirement plans for seniors.

\begin{itemize}
    \item \textbf{Data Collection}: Age data is often collected during user registration processes, through user-provided information, or inferred from behavior and interests (e.g., browsing history, content consumption).
    \item \textbf{Indexing}: Ads targeting specific age groups are indexed with age-related keywords. For example, an ad for a college savings plan might be indexed with keywords related to young adults and parents.
    \item \textbf{Matching}: The inverted index matches the user's age profile with relevant ads, ensuring that users see advertisements tailored to their age group.
\end{itemize}
\subsection{Gender Targeting}
Gender targeting aims to deliver ads based on the user's gender, a crucial demographic factor for many products and services.

\begin{itemize}
    \item \textbf{Data Collection}: Gender information is collected through user profiles, social media interactions, and inferred from behavior (e.g., types of products viewed or purchased).
    \item \textbf{Indexing}: Ads are indexed with gender-related keywords. For instance, an ad for men's grooming products would be indexed with keywords related to male users.
    \item \textbf{Matching}: The system matches the user's gender profile with relevant ads, ensuring that users receive advertisements that are likely to be more appealing to their gender.
\end{itemize}

\subsection{Re-targeting}
Re-targeting, also known as remarketing, focuses on users who have previously interacted with a website or app but did not complete a desired action, such as making a purchase.
\begin{figure}[h]
    \centering
    \includegraphics[width=\linewidth]{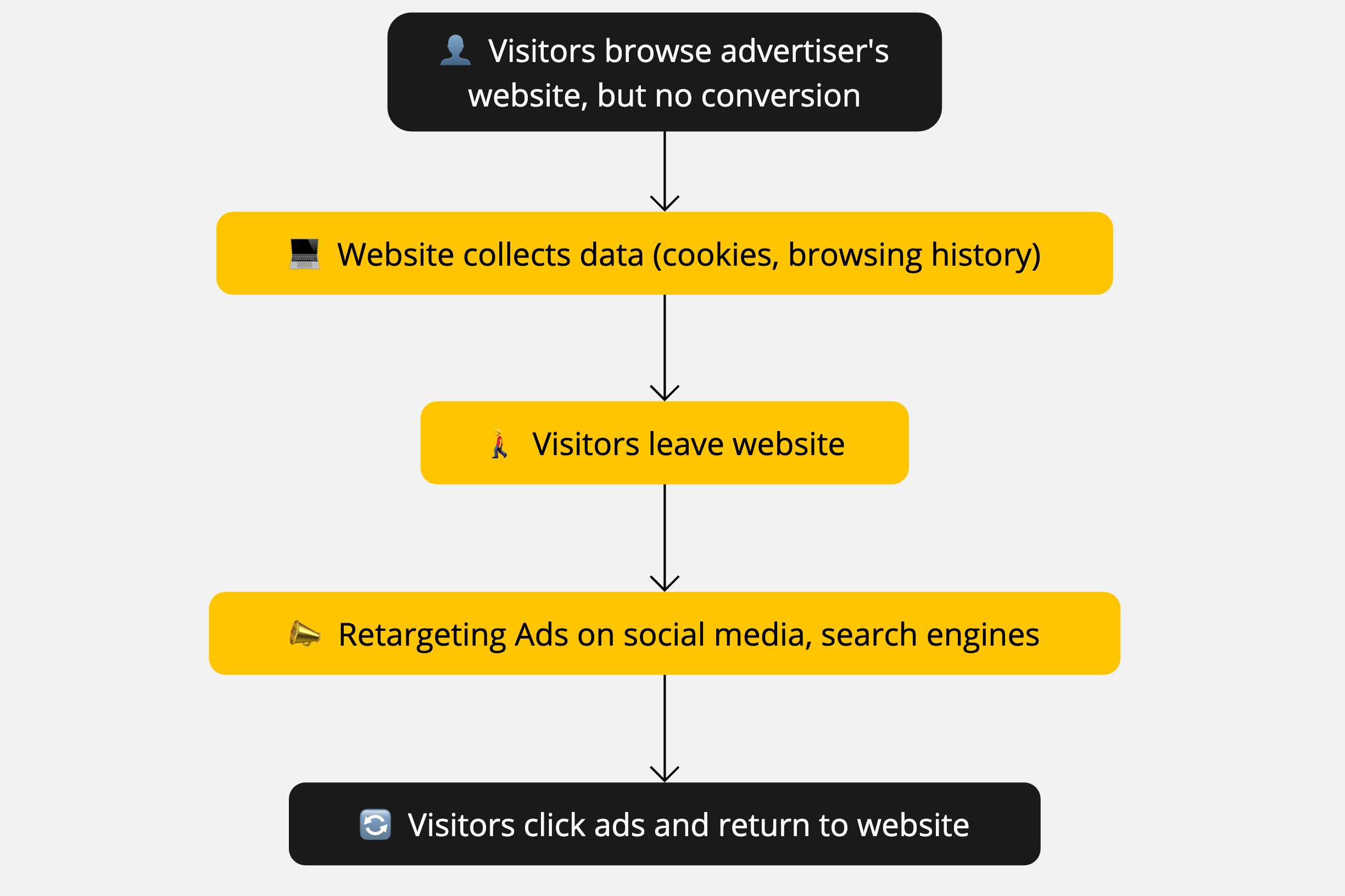}
    \caption{Re-targeting}
    \label{fig:Re-targeting}
\end{figure}

\begin{itemize}
    \item \textbf{Data Collection}: User interaction data is collected through cookies, tracking pixels, and other tracking technologies. This data includes pages visited, products viewed, and actions taken.
    \item \textbf{Indexing}: Ads are indexed based on the specific actions users took. For example, an ad for a product left in a shopping cart might be indexed with keywords related to that product and the action of adding to the cart.
    \item \textbf{Matching}: When a user who has been re-targeted returns to the site or browses the web, the system matches their previous interactions with relevant ads, reminding them of the products or services they showed interest in.
\end{itemize}
\subsection{Keyword Targeting}
Keyword targeting involves delivering ads based on specific keywords that users have entered in search queries or content they are currently viewing.

Large language models (LLMs) can significantly enhance keyword targeting by rewriting and expanding keywords, thereby improving the precision and reach of ad campaigns. LLMs like GPT-4 can generate a diverse set of keyword variations that capture different ways users might express their search intent. For example, an LLM can transform a single keyword like “running shoes” into multiple related phrases such as “athletic footwear,” “jogging sneakers,” and “marathon trainers.” This keyword expansion allows advertisers to cover a broader range of search queries, increasing the likelihood of matching user intent and improving ad relevance. Additionally, LLMs can refine and optimize keywords based on contextual understanding, ensuring that the generated keywords align with the overall theme and target audience of the ad campaign. \cite{zhaoutilizing} applied LLM in this fashion to analyze legal documents and produce numerical features from text data. By leveraging LLMs for keyword rewriting, advertisers can enhance their targeting strategies, drive more relevant traffic, and ultimately achieve better campaign performance.
\begin{itemize}
    \item \textbf{Data Collection}: Keywords are collected from search queries, page content, and user-generated content such as comments and reviews.
    \item \textbf{Indexing}: Ads are indexed with relevant keywords. For example, an ad for "running shoes" would be indexed with keywords such as "running," "shoes," "fitness," and "sports."
    \item \textbf{Matching}: The system matches the keywords associated with the user's current activity or interests with relevant ads, ensuring that users see advertisements related to their immediate context or search intent.
\end{itemize}
\subsection{Behavioral Targeting}
Behavioral targeting is a key technique used in ad recommendation systems. It involves tracking users' online behavior to deliver personalized ads.

\begin{itemize}
    \item \textbf{Data Collection}: Behavioral data is collected through user activities such as browsing history, search queries, social media interactions, and purchase behavior. This data is often aggregated over time to build comprehensive user profiles.
    \item \textbf{Indexing}: Ads are indexed based on behavioral patterns. For example, an ad for outdoor equipment might be indexed with keywords related to activities like hiking, camping, and adventure sports.
    \item \textbf{Matching}: The system analyzes the user's behavior and matches it with relevant ads. For instance, if a user frequently searches for hiking gear, the system will show ads for outdoor equipment. This method increases the likelihood of ad engagement by presenting users with ads that match their interests and needs.
\end{itemize}

Behavioral targeting improves ad relevance and effectiveness by focusing on users' demonstrated interests and preferences rather than relying solely on static demographic data.

\subsection{Contextual Targeting}
Contextual targeting involves displaying ads based on the content of the webpage a user is currently viewing.

\begin{itemize}
    \item \textbf{Data Collection}: Contextual data is collected from the content of the webpage, including text, images, and metadata. Natural language processing (NLP) and other content analysis techniques are used to understand the context.
    \item \textbf{Indexing}: Ads are indexed with keywords related to specific contexts. For example, an ad for gym memberships might be indexed with keywords related to fitness, health, and exercise.
    \item \textbf{Matching}: The system matches the content of the webpage with relevant ads. For instance, an article about fitness might feature ads for gym memberships or sportswear. This method ensures that ads are relevant to the user's immediate context, enhancing their effectiveness.
\end{itemize}

Contextual targeting is particularly effective for capturing user attention when they are already engaged with related content, increasing the chances of ad interaction and conversion.

\section{Retrieval Systems}
Ad targeting focuses on delivering personalized advertisements to users based on their profiles and behaviors, aiming to maximize engagement and conversion rates. Organic retrieval, on the other hand, emphasizes improving user experience by recommending content or products that align with users' preferences without any direct monetary influence. 

\subsection{Organic Retrieval}

Organic retrieval, also known as organic recommendation systems, aims to enhance user experience by providing personalized content or product recommendations without direct monetary influence. These systems analyze user data to suggest items that align with users' preferences and behaviors.

Organic retrieval systems are used in various domains, including e-commerce, streaming services, and social media platforms. For example, Netflix's recommendation system suggests movies and shows that align with users' viewing habits, while Spotify's Discover Weekly playlist introduces users to new music based on their listening history.

Organic retrieval systems face several challenges, including data sparsity, the cold start problem, and maintaining diversity in recommendations. Data sparsity occurs when there is insufficient data to make accurate recommendations, while the cold start problem arises with new users or items that lack historical data. Maintaining diversity is crucial to prevent the system from becoming too narrow in its suggestions, which can lead to user fatigue.

\subsubsection{Mechanisms of Organic Retrieval}

\begin{itemize}
\item \textbf{Content-Based Filtering}

Content-based filtering focuses on the characteristics of items. For instance, if a user has shown interest in science fiction books, the system will recommend other books in the same genre.

\item \textbf{Collaborative Filtering}

Collaborative filtering makes recommendations based on the preferences of similar users. It identifies patterns and correlations among users’ behaviors and suggests items that users with similar tastes have liked. There are two main types of collaborative filtering: user-based and item-based.

\item \textbf{Hybrid Systems}

Many modern recommendation systems use a hybrid approach, combining both content-based and collaborative filtering techniques to improve accuracy and relevance. These systems can provide more nuanced recommendations by leveraging the strengths of both methods.
\end{itemize}

\subsection{Two-Tower Network for Retrieval}
The two-tower model, also known as the dual-tower model, is a deep learning architecture widely used in recommendation systems for content retrieval. It expands the matrix factorization model \cite{koenigstein2012efficient} by introducing diverse types of features into the model. It consists of two separate neural networks (towers): one for encoding user features and the other for encoding item features. The key idea behind this model is to project users and items into a shared latent space where their compatibility can be measured. \cite{agarwal2012fast} provides a introductory explanation on top-k retrieval using model based recommendation. 

\subsubsection{Architecture of the Two-Tower Model}

The two-tower model comprises two main components:

\begin{figure}[h]
    \centering
    \includegraphics[width=\linewidth]{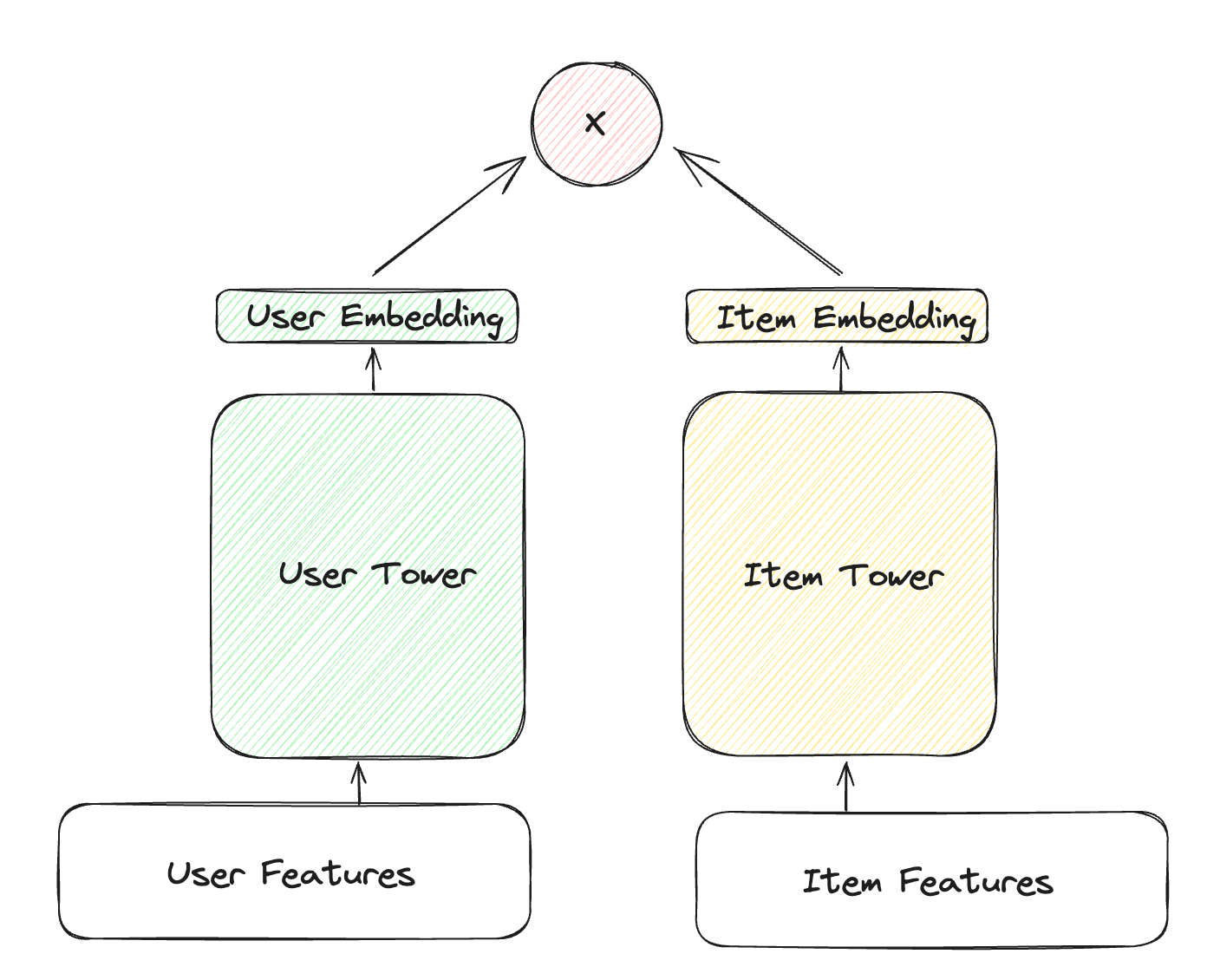}
    \caption{Two-Tower Neural Network}
    \label{fig:two_tower}
\end{figure}

\subsubsection{User Tower}

The user tower is designed to capture and encode user features. These features can include demographic information, browsing history, past interactions, and explicit preferences. The user tower typically consists of multiple layers of neural networks that transform the raw input features into a dense vector representation in the latent space. The architecture may include:

\begin{itemize}
    \item \textbf{Embedding Layers}: These layers are used to convert categorical features (e.g., user ID, gender, age group) into dense vectors.
    \item \textbf{Fully Connected Layers}: These layers perform non-linear transformations on the embedded features, capturing complex relationships and interactions between different user attributes.
    \item \textbf{Normalization and Regularization}: Techniques such as batch normalization and dropout are employed to improve the model's generalization ability and prevent overfitting.
\end{itemize}

\subsubsection{Item Tower}

The item tower operates similarly to the user tower but focuses on encoding item features. These features can include item metadata (e.g., title, category, tags), content characteristics (e.g., text, images), and contextual information (e.g., time of interaction, device type). The item tower also transforms the raw input features into a dense vector representation in the same latent space as the user tower. The architecture may include:

\begin{itemize}
    \item \textbf{Embedding Layers}: These layers convert categorical item features into dense vectors.
    \item \textbf{Feature Extractors}: For content-rich items (e.g., text or images), specialized layers such as convolutional neural networks (CNNs) or recurrent neural networks (RNNs) may be used to extract meaningful features.
    \item \textbf{Fully Connected Layers}: These layers transform the embedded features into dense vectors.
    \item \textbf{Normalization and Regularization}: Similar techniques as in the user tower are applied to enhance the model's performance.
\end{itemize}

\subsection{Training the Two-Tower Model}

Training the two-tower model involves optimizing the latent representations such that the distance or similarity between user and item vectors accurately reflects their compatibility. The following steps outline the training process:

\subsubsection{Pairing Users and Items}

During training, pairs of users and items are sampled from historical interaction data. Positive pairs consist of users and items with known interactions (e.g., a user watched a video), while negative pairs are sampled to represent non-interactions (e.g., a user did not watch a video). This sampling strategy helps the model learn to distinguish between relevant and irrelevant items for each user. Optimizations such as \cite{yang2020mixed} can be applied to address the imbalance between positive and negative examples. 

\subsubsection{Loss Function}

A common loss function used in training the two-tower model is the cross-entropy loss or the pairwise ranking loss. These loss functions measure the model's ability to correctly predict the relevance of items for users. The pairwise ranking loss, for example, aims to ensure that the score for positive pairs is higher than that for negative pairs by a certain margin.

\subsubsection{Optimization}

The model parameters are optimized using gradient-based optimization algorithms such as stochastic gradient descent (SGD) or Adam. During each training iteration, the gradients of the loss function with respect to the model parameters are computed and used to update the weights of the neural networks.

When attention-based architectures, such as Transformers, are employed within each tower of the two-tower model, the training process demands significantly more computational resources. Attention mechanisms enable the model to focus on relevant parts of the input data, improving its ability to capture intricate relationships between users and items. However, this comes at the cost of increased computational complexity, as each attention layer involves numerous matrix multiplications and operations that scale quadratically with the input size. \cite{mei2024efficiency} have highlighted this efficiency bottleneck, pointing out that the heavy computational requirements and memory usage of attention-based architectures can slow down the training process and limit scalability. Efficiently training these models necessitates advanced optimization techniques, high-performance hardware, and potentially distributed computing strategies to manage the increased load and ensure that the model converges within a reasonable timeframe.

\subsection{Two-Tower Inference and Retrieval}

Once the two-tower model is trained, it can be used for real-time content retrieval. The inference process involves:

\subsubsection{User and Item Encoding}

For a given user, the user tower generates a dense vector representation based on the user's features. Similarly, the item tower generates dense vector representations for all candidate items.

\subsubsection{Similarity Computation}

The similarity between the user vector and each item vector is computed using a similarity metric such as the dot product or cosine similarity. This step ranks the items based on their relevance to the user.

\subsubsection{Top-N Recommendation}

The top-N items with the highest similarity scores are selected and recommended to the user. This approach ensures that the recommendations are tailored to the user's preferences and behaviors.

\subsection{Advantages of the Two-Tower Model}

The two-tower model offers several advantages for organic content retrieval:

\subsubsection{Scalability}

The decoupled architecture of the two-tower model allows for efficient scaling. Since the user and item towers are independent, new users and items can be added without retraining the entire model. This scalability is particularly beneficial for platforms with large and dynamic content libraries.

\subsubsection{Flexibility}

The model's architecture is highly flexible and can accommodate various types of features and content modalities. Whether the items are text-based (e.g., articles), image-based (e.g., photos), or a combination of different media, the two-tower model can be adapted to encode and retrieve them effectively.

\subsubsection{Personalization}

By learning to represent users and items in a shared latent space, the two-tower model can capture complex and nuanced relationships between user preferences and item characteristics. This capability enables highly personalized recommendations that align with individual user interests.

\subsubsection{Real-Time Retrieval}

The model's ability to compute similarity scores between user and item vectors in real time makes it well-suited for applications requiring immediate responses, such as online recommendation systems. The precomputed item vectors can be efficiently indexed and searched, facilitating quick retrieval.

\subsection{Challenges and Considerations}

While the two-tower model is powerful, it also presents certain challenges and considerations:

\subsubsection{Cold Start Problem}

The cold start problem refers to the difficulty of making accurate recommendations for new users or items with little or no historical interaction data. Addressing this issue may require incorporating additional features or leveraging transfer learning techniques. \cite{xin2021atnn} applied multiple learning scheme to two-tower neural networks to address the cold start problem. 

\subsubsection{Data Quality}

The effectiveness of the two-tower model relies heavily on the quality and quantity of the input data. Inaccurate or sparse data can lead to poor model performance and suboptimal recommendations. Ensuring high-quality data collection and preprocessing is crucial.

\subsubsection{Possible Improvements to Two-Tower Model }
Improving the traditional two-tower model can be achieved through several innovative approaches, such as implementing a multi-task two-tower model \cite{zhang2018overview} and developing a three-tower architecture.

The multi-task two-tower model enhances the original framework by enabling it to handle multiple tasks simultaneously. This is accomplished by sharing the representation space between different tasks, allowing the model to learn from varied data sources and contexts. For instance, in recommendation systems, this approach can combine tasks like click-through rate prediction and user engagement metrics, leading to more robust and generalized user embeddings. By leveraging shared representations, the multi-task model not only improves prediction accuracy but also enhances the model's ability to generalize across different tasks, ultimately providing a more comprehensive understanding of user preferences and behaviors.

The three-tower model introduces an additional tower to the traditional architecture, aimed at integrating more diverse information sources or auxiliary data. In the context of a recommendation system, while the original two towers might represent user and item embeddings, the third tower could incorporate contextual information such as temporal dynamics or location-based data. This extra tower helps to capture more complex interactions and dependencies that are not easily modeled with just two towers. By enriching the model with additional dimensions of data, the three-tower architecture can significantly improve the accuracy and relevance of recommendations. Furthermore, this approach allows for more fine-grained personalization by considering a broader spectrum of influencing factors, thereby enhancing the overall user experience and satisfaction.

\begin{figure}[h]
    \centering
    \includegraphics[width=\linewidth]{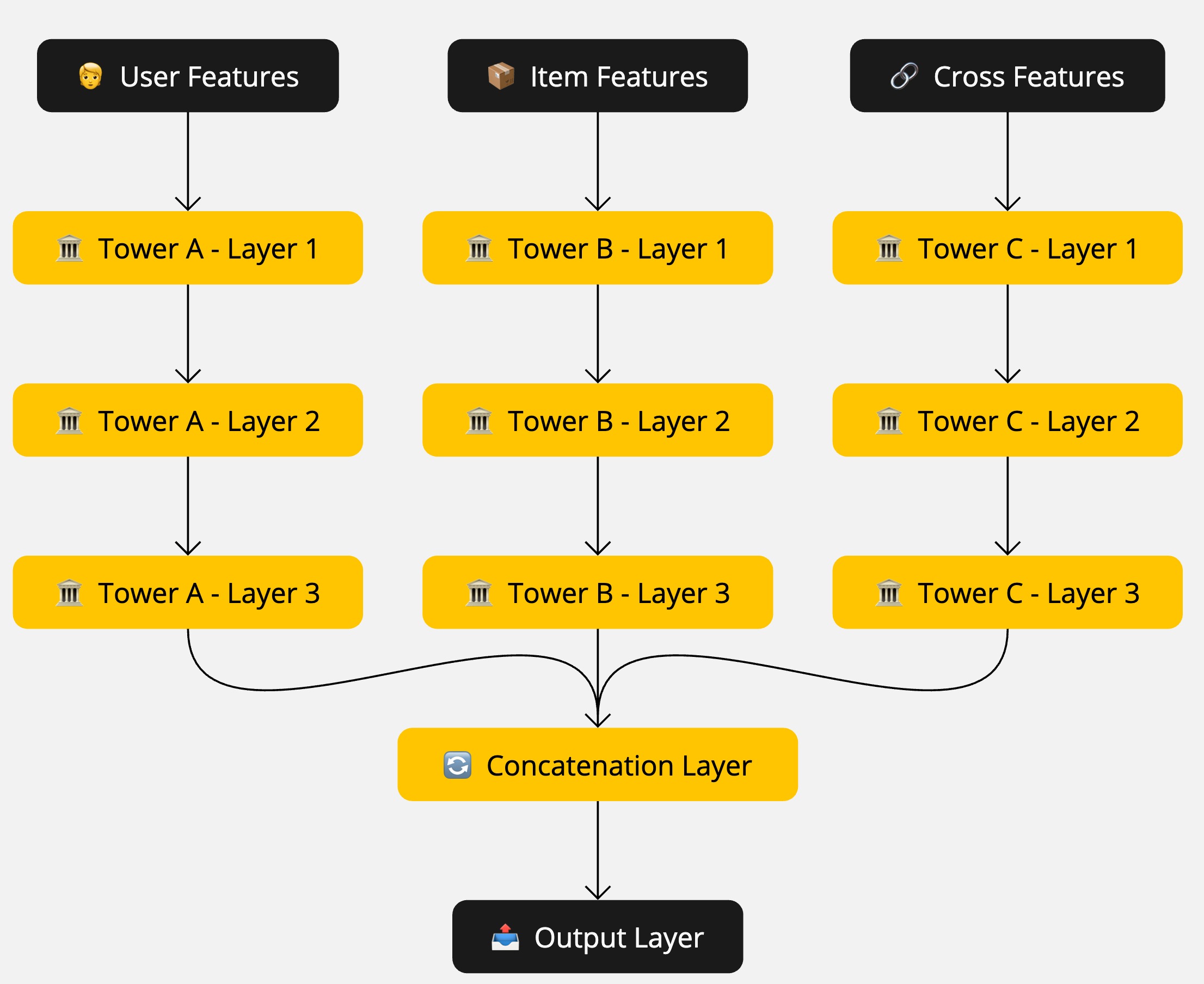}
    \caption{Three Tower Neural Network}
    \label{fig:three_tower}
\end{figure}

Another potential improvement to the two-tower model for recommendation systems involves leveraging large language models (LLMs) to expand datasets, enhancing the model’s robustness and accuracy. LLMs, such as GPT-4 \cite{achiam2023gpt}, can generate synthetic data that mimics user interactions, item descriptions, and contextual information, thereby enriching the training dataset \cite{li2024feature}. This can be particularly beneficial in addressing the cold start problem, where new users or items lack sufficient historical data. By generating realistic and diverse synthetic interactions, LLMs can help the two-tower model learn more comprehensive patterns and relationships, improving its ability to make accurate recommendations \cite{zhao2024utilizing} \cite{yangusing}. Additionally, LLMs can enhance the semantic understanding of textual features, providing richer embeddings for both users and items. This expanded and enriched dataset allows the two-tower model to generalize better across different user behaviors and item characteristics, ultimately leading to more personalized and relevant recommendations. In addition to LLMs, traditional machine learning models can be used to simulate datasets with particular characteristics. \cite{Shen2024Harnessing} applied XGBoost to simulate datasets to mimic correlation structures commonly found in imaging data. 

\section{Comparing Ad Targeting and Organic Retrieval}
On a high level, ad targeting shares many similarities with organic retrieval, as both processes are fundamentally about matching user intent with the most relevant content. In ad targeting, algorithms analyze vast amounts of user data to predict and deliver ads that are most likely to engage the user, just as search engines analyze web content to rank and display the most relevant search results. Both systems rely heavily on data processing, pattern recognition, and machine learning to continually refine their accuracy and effectiveness. Whether through analyzing search queries and web content or user behavior and demographic information, the ultimate goal is to create a seamless and personalized user experience by presenting the most pertinent information, be it an advertisement or an organic search result. This convergence in methodology underscores the growing intersection between targeted advertising and content retrieval technologies. 

Index-based ad targeting is similar to the inverted index commonly used in organic retrievals. Both techniques rely on the creation and use of a structured index to efficiently locate and serve relevant information. In ad targeting, advertisers build indices that categorize users based on their behaviors, interests, and demographics. This allows for precise targeting of ads to users who are most likely to be interested in the advertised product or service. Similarly, in organic retrieval, recommendation systems and search engines use an inverted index to map content to keys. This index allows the search engine to quickly retrieve and rank relevant documents in response to a user’s query, ensuring that the most pertinent content appears in search results.

Targeting options such as interest targeting and keyword targeting in advertising are akin to the different channels used to retrieve content in organic searches. Interest targeting involves showing ads to users based on their demonstrated interests and behaviors online, much like how a search engine retrieves content related to a user’s query by analyzing the context and relevance of indexed pages. Keyword targeting, on the other hand, focuses on specific words or phrases, similar to how search engines match user queries with the indexed content that contains those keywords. Both advertising and organic retrieval systems aim to connect users with content that aligns with their interests or search intents, optimizing the relevance and effectiveness of the information delivered.

Next, we will outline some key differences between the two retrieval systems.
\subsubsection{Objectives}

The primary objective of ad targeting is to drive revenue by delivering targeted advertisements to users. The focus is on optimizing ad placements to increase click-through rates and conversions, thereby generating revenue for both the platform and advertisers.

In contrast, the primary goal of organic retrieval is to enhance user satisfaction and retention by providing relevant and engaging content. These systems aim to improve user experience and foster long-term loyalty by helping users discover new items that align with their preferences.

\subsubsection{Data Utilization}

Ad targeting relies heavily on detailed user profiling and behavioral data. This includes demographic information, browsing history, search queries, social media activity, and purchase behavior. The aim is to build comprehensive user profiles that can be used to deliver highly personalized ads.

Organic retrieval systems also use user behavior data, but the focus is on historical interactions and preferences rather than demographic profiling. These systems analyze users' past behaviors to make personalized content or product recommendations.

\subsubsection{Impact on User Experience}

Ad targeting can significantly impact user experience by delivering highly relevant ads that match users' interests and needs. However, it can also lead to ad fatigue if users are bombarded with too many ads. Additionally, privacy concerns arise as users become more aware of how their data is being collected and used.

Organic retrieval systems aim to enhance user experience by providing personalized content or product recommendations. These systems help users discover new items and engage more deeply with the platform. However, challenges such as data sparsity and the cold start problem can affect the accuracy and relevance of recommendations.

\subsubsection{Revenue Generation}

Ad targeting is primarily focused on revenue generation. By delivering targeted ads, platforms can charge advertisers higher rates for ad placements, resulting in increased revenue. Real-time bidding further maximizes ad revenue by selling ad impressions to the highest bidder.

Organic retrieval systems do not directly generate revenue through ads. Instead, they enhance user engagement and retention, which can indirectly lead to increased revenue. For example, a streaming service with an effective recommendation system may see higher subscription renewal rates due to improved user satisfaction.

\subsubsection{Privacy and Ethical Concerns}

Ad targeting raises significant privacy concerns as it involves extensive data collection and profiling. Users may feel uncomfortable with the level of personal information being tracked and used for advertising purposes. There are also ethical concerns regarding the manipulation of user behavior through targeted ads \cite{goldfarb2011online}.

Organic retrieval systems also collect and analyze user data, but the focus is on improving user experience rather than monetizing user profiles. While privacy concerns still exist, they are generally less pronounced compared to ads.

\section{Metrics and Experimentation}
\subsection{Metrics}
Content recommendation systems primarily focus on metrics that reflect user engagement and retention, such as Monthly Active Users (MAU), Daily Active Users (DAU), and user retention rates. These systems aim to enhance the user experience by delivering personalized content that aligns with individual preferences, thereby encouraging more frequent and prolonged interactions with the platform. The ultimate goal is to increase user loyalty and satisfaction, leading to sustained growth in user base and activity levels.

In addition to these high-level metrics, more granular metrics like clicks, likes, and follows are also crucial. These metrics are particularly sensitive to algorithmic updates and provide immediate feedback on user engagement with the recommended content. They are closely monitored to gauge the effectiveness of the recommendation algorithms and to make timely adjustments that can further enhance user interaction and satisfaction.

In contrast, ad systems operate within a three-party ecosystem involving advertisers, the platform, and users. Each party has distinct interests and metrics for success. Advertisers are primarily concerned with metrics such as Cost Per Click (CPC) and conversion rates, which indicate the effectiveness and efficiency of their ad spend. The platform, on the other hand, measures revenue generated from ad placements, aiming to maximize overall profitability. Users’ interests lie in the relevance and non-intrusiveness of the ads they encounter; high relevance can enhance user experience, while irrelevant or excessive ads can detract from it. Balancing these three sets of interests is crucial for the success of ad systems.

\subsection{Controlled Experimentation Framework}

To evaluate and optimize both content recommendation and ad systems, controlled A/B experimentation frameworks are commonly used \cite{kohavi2015online}. In this setup, users are randomly divided into control and treatment groups. The control group continues to experience the system as usual, while the treatment group is exposed to a new feature or change being tested. By comparing the behavior and outcomes of these two groups, the impact of the new feature can be accurately assessed. \cite{knijnenburg2015evaluating}

\begin{figure}[h]
    \centering
    \includegraphics[width=\linewidth]{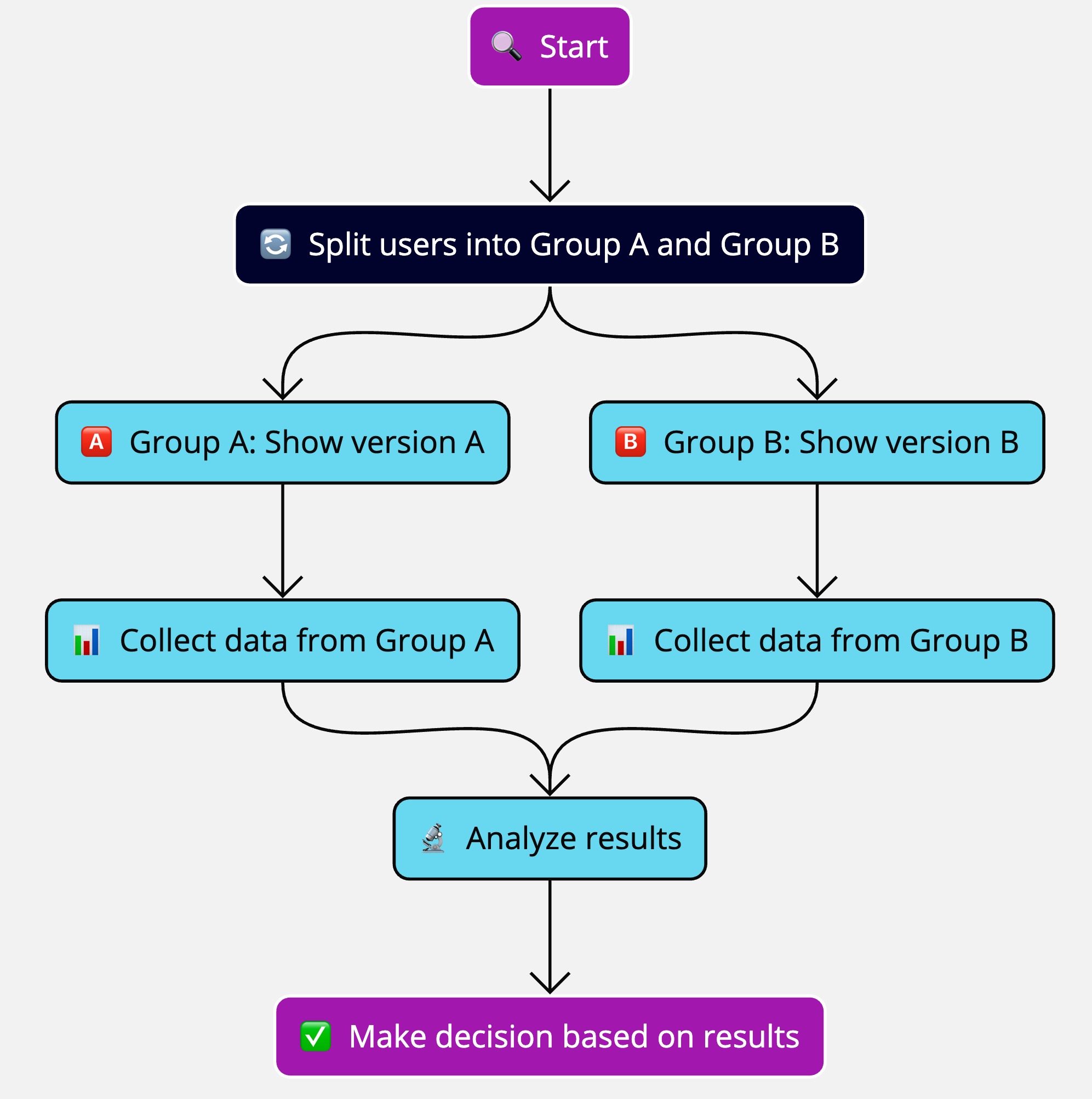}
    \caption{AB Experimentation}
    \label{fig:Experimentation}
\end{figure}

However, A/B experiments in these systems come with unique challenges. For user experiments in content recommendation, there is a risk of traffic stealing, where the treatment group may inadvertently attract users who would otherwise belong to the control group. This can skew the results, making it difficult to measure the true effect of the experiment. Metrics such as clicks, likes, and follows are particularly sensitive to algorithmic updates and need to be closely monitored during these experiments to assess the immediate impact on user engagement.

When experimenting with content creators or advertisers, it’s essential to separate users and creators to avoid “ghost experimentation” differences. This means ensuring that users and content creators (or advertisers) do not cross-contaminate the control and treatment groups. For instance, if an experiment is run with a subset of content creators, the audience for these creators should be evenly distributed between control and treatment groups to ensure accurate measurement of the experiment’s impact.

For ad experiments, it is also critical to ensure that both the treatment and control groups have the same budget. This helps to maintain a fair comparison between the two groups and ensures that any differences in performance can be attributed to the experimental changes rather than budget discrepancies.

\section{Conclusion}
The landscape of retrieval algorithms in ad recommendation and content recommendation systems is vast and continually evolving. This survey has highlighted the critical role these algorithms play in enhancing user engagement, driving revenue, and improving overall user satisfaction. The two-tower model and its emerging variants, such as the multi-task and three-tower models, demonstrate the potential for increased accuracy and personalization in recommendations. However, the deployment of these systems comes with significant challenges, including data quality, privacy concerns, and the cold start problem.

As digital platforms continue to expand and user expectations grow, the development of more sophisticated and ethical retrieval algorithms will be essential. Future research should focus on addressing these challenges, ensuring that recommendation systems not only provide relevant and engaging content but also uphold user privacy and data integrity. By striking a balance between personalization and ethical considerations, the next generation of retrieval algorithms can offer even greater value to both users and businesses.

\bibliographystyle{named}
\bibliography{ms}

\begin{thebibliography}{}

\bibitem[\protect\citeauthoryear{Achiam \bgroup \em et al.\egroup }{2023}]{achiam2023gpt}
Josh Achiam, Steven Adler, Sandhini Agarwal, Lama Ahmad, Ilge Akkaya, Florencia~Leoni Aleman, Diogo Almeida, Janko Altenschmidt, Sam Altman, Shyamal Anadkat, et~al.
\newblock Gpt-4 technical report.
\newblock {\em arXiv preprint arXiv:2303.08774}, 2023.

\bibitem[\protect\citeauthoryear{Agarwal and Gurevich}{2012}]{agarwal2012fast}
Deepak Agarwal and Maxim Gurevich.
\newblock Fast top-k retrieval for model based recommendation.
\newblock In {\em Proceedings of the fifth ACM international conference on Web search and data mining}, pages 483--492, 2012.

\bibitem[\protect\citeauthoryear{Goldfarb and Tucker}{2011}]{goldfarb2011online}
Avi Goldfarb and Catherine Tucker.
\newblock Online display advertising: Targeting and obtrusiveness.
\newblock {\em Marketing Science}, 30(3):389--404, 2011.

\bibitem[\protect\citeauthoryear{Hu and Lu}{2024}]{hu2024rag}
Yucheng Hu and Yuxing Lu.
\newblock Rag and rau: A survey on retrieval-augmented language model in natural language processing.
\newblock {\em arXiv preprint arXiv:2404.19543}, 2024.

\bibitem[\protect\citeauthoryear{Knijnenburg and Willemsen}{2015}]{knijnenburg2015evaluating}
Bart~P Knijnenburg and Martijn~C Willemsen.
\newblock Evaluating recommender systems with user experiments.
\newblock In {\em Recommender systems handbook}, pages 309--352. Springer, 2015.

\bibitem[\protect\citeauthoryear{Koenigstein \bgroup \em et al.\egroup }{2012}]{koenigstein2012efficient}
Noam Koenigstein, Parikshit Ram, and Yuval Shavitt.
\newblock Efficient retrieval of recommendations in a matrix factorization framework.
\newblock In {\em Proceedings of the 21st ACM international conference on Information and knowledge management}, pages 535--544, 2012.

\bibitem[\protect\citeauthoryear{Kohavi and Longbotham}{2015}]{kohavi2015online}
Ron Kohavi and Roger Longbotham.
\newblock Online controlled experiments and a/b tests.
\newblock {\em Encyclopedia of machine learning and data mining}, pages 1--11, 2015.

\bibitem[\protect\citeauthoryear{Li \bgroup \em et al.\egroup }{2024}]{li2024feature}
Zhenglin Li, Yangchen Huang, Mengran Zhu, Jingyu Zhang, JingHao Chang, and Houze Liu.
\newblock Feature manipulation for ddpm based change detection.
\newblock {\em arXiv preprint arXiv:2403.15943}, 2024.

\bibitem[\protect\citeauthoryear{Mei \bgroup \em et al.\egroup }{2024}]{mei2024efficiency}
Taiyuan Mei, Yun Zi, Xiaohan Cheng, Zijun Gao, Qi~Wang, and Haowei Yang.
\newblock Efficiency optimization of large-scale language models based on deep learning in natural language processing tasks.
\newblock {\em arXiv preprint arXiv:2405.11704}, 2024.

\bibitem[\protect\citeauthoryear{Scholer \bgroup \em et al.\egroup }{2002}]{scholer2002compression}
Falk Scholer, Hugh~E Williams, John Yiannis, and Justin Zobel.
\newblock Compression of inverted indexes for fast query evaluation.
\newblock In {\em Proceedings of the 25th annual international ACM SIGIR conference on Research and development in information retrieval}, pages 222--229, 2002.

\bibitem[\protect\citeauthoryear{Shen \bgroup \em et al.\egroup }{2024}]{Shen2024Harnessing}
Xinyu Shen, Qimin Zhang, Huili Zheng, and Weiwei Qi.
\newblock Harnessing xgboost for robust biomarker selection of obsessive-compulsive disorder (ocd) from adolescent brain cognitive development (abcd) data.
\newblock {\em ResearchGate}, May 2024.

\bibitem[\protect\citeauthoryear{Xin \bgroup \em et al.\egroup }{2021}]{xin2021atnn}
SHEN Xin, Zhao Li, Pengcheng Zou, Cheng Long, Jie Zhang, Jiajun Bu, and Jingren Zhou.
\newblock Atnn: adversarial two-tower neural network for new item’s popularity prediction in e-commerce.
\newblock In {\em 2021 IEEE 37th International Conference on Data Engineering (ICDE)}, pages 2499--2510. IEEE, 2021.

\bibitem[\protect\citeauthoryear{Yan \bgroup \em et al.\egroup }{2009}]{yan2009much}
Jun Yan, Ning Liu, Gang Wang, Wen Zhang, Yun Jiang, and Zheng Chen.
\newblock How much can behavioral targeting help online advertising?
\newblock In {\em Proceedings of the 18th international conference on World wide web}, pages 261--270, 2009.

\bibitem[\protect\citeauthoryear{Yang \bgroup \em et al.\egroup }{2020}]{yang2020mixed}
Ji~Yang, Xinyang Yi, Derek Zhiyuan~Cheng, Lichan Hong, Yang Li, Simon Xiaoming~Wang, Taibai Xu, and Ed~H Chi.
\newblock Mixed negative sampling for learning two-tower neural networks in recommendations.
\newblock In {\em Companion proceedings of the web conference 2020}, pages 441--447, 2020.

\bibitem[\protect\citeauthoryear{Yang \bgroup \em et al.\egroup }{2024}]{yangusing}
Shiqi Yang, Yu~Zhao, and Haoxiang Gao.
\newblock Using large language models in real estate transactions: A few-shot learning approach.
\newblock {\em OSF Preprints}, May 2024.

\bibitem[\protect\citeauthoryear{Zhang and Yang}{2018}]{zhang2018overview}
Yu~Zhang and Qiang Yang.
\newblock An overview of multi-task learning.
\newblock {\em National Science Review}, 5(1):30--43, 2018.

\bibitem[\protect\citeauthoryear{Zhao and Gao}{2024}]{zhao2024utilizing}
Yu~Zhao and Haoxiang Gao.
\newblock Utilizing large language models for information extraction from real estate transactions.
\newblock {\em arXiv preprint arXiv:2404.18043}, 2024.

\bibitem[\protect\citeauthoryear{Zhao \bgroup \em et al.\egroup }{2024}]{zhaoutilizing}
Yu~Zhao, Haoxiang Gao, and Shiqi Yang.
\newblock Utilizing large language models to analyze common law contract formation.
\newblock {\em OSF Preprints}, Jun 2024.

\end{thebibliography}

\end{document}